\documentclass[prb,showpacs,floatfix,twocolumn]{revtex4}
\usepackage{graphicx}
\usepackage{amssymb}
\usepackage{dcolumn}
\usepackage{bm}
\begin{document}

\title{Velocity weakening and possibility of aftershocks in nanofriction experiments} 
\author{E. A. Jagla}

\affiliation{Centro At\'omico Bariloche and Instituto Balseiro, Comisi\'on Nacional de Energ\'{\i}a At\'omica, 
(8400) Bariloche, Argentina}

\begin{abstract}

We study the frictional behavior of small contacts as those realized in the atomic force microscope and other experimental setups, in the framework of generalized Prandtl-Tomlinson models. 
Particular attention is paid to mechanisms that generate velocity weakening, namely a decreasing average friction force with the relative sliding velocity.
The mechanisms studied model the possibility of viscous relaxation, or aging effects in the contact. 
It is found that, in addition to producing velocity weakening, these mechanisms can also produce aftershocks at sufficiently low sliding velocities. This provides a remarkable analogy at the microscale, of friction properties at the macroscale, where aftershocks and velocity weakening are two fundamental features of seismic phenomena.

\end{abstract}
\maketitle
\newpage

\section{Introduction}

In the last years, experimental achievements have given the possibility to study friction at the microscale (nanofriction),
which has renewed the interest in a variety of theoretical and applied aspects of tribology. The prototypical experimental
setup is that of the atomic force microscope (AFM) where a nanometer size tip can be brought into contact with a substrate, and the frictional behavior of the system can be studied in detail \cite{carpick0}. Other related instruments, like the surface force apparatus\cite{sfa} have expanded the possibilities to cover other regions of parameters, allowing for instance to study somewhat larger contact areas, in the range of microns. 

The properties of nanofriction differ substantially from those at a macroscopic scale. One reason for this is that macroscopic friction originates typically in a set of sparse contacts located at different spatial positions, and the observed behavior is a non-trivial combination of these individual contacts. The now available experimental techniques give the opportunity to study one such individual contact in detail. 

One not fully attained goal in the field of friction is precisely to correlate the results of well controlled nanofriction experiments with those of friction at a macroscopic scale. For instance, in the largest scale realization available, namely that corresponding to seismic phenomena, friction displays features that seem to be typical of this large scale. One of them is velocity weakening (VW). This refers to a friction process in which the average friction force is a decreasing function of the relative velocity of the sliding bodies. It is widely accepted that VW is a necessary condition for seismic phenomena to look the way they do\cite{scholz}. 
Although on a phenomenological perspective velocity weakening is successfully modeled by the so-called rate and state equations\cite{dietrich,ruina}, there is no general agreement on a consistent first principle description of the phenomenon.
It is however agreed upon the fact that VW must be induced by some kind of mechanism that tend to strengthen the contact between the two sliding bodies if they remain in the same relative position for a long time.

Probably the most typical feature of friction at the seismic scale is the existence of aftershocks (ASs). They occur as secondary shocks after some large earthquake. ASs occur exclusively due to an internal dynamics of the sliding bodies, and are not directly related to the driving mechanism. ASs have been shown to be compatible with rate and state equations displaying VW\cite{rs_and_as}. In addition,  in recently studied numerical models of seismicity, internal relaxation on the plates have shown to give rise at the same time to VW and ASs.\cite{jagla1,jagla2}

The dependence of friction force on the relative velocity between the two sliding objects is routinely measured in nanofriction experiments. In many cases, this dependence follows an increasing dependence on velocity. In a minority of cases, VW is observed.
VW is well established at the scale probed by the surface force apparatus, where it has been directly measured\cite{marone,marone2}.
Single contact AFM measurements have sometimes also found a VW friction law. 

The physical mechanisms that have been proposed as generating VW in nanofriction experiments typically include the existence of some internal dynamics in the contact that strengthens the tip-substrate interaction as a function of time. For instance, in \cite{riedo} it was argued that in the AFM, capillary condensation of water molecules occur at the contact between tip and substrate. 
The VW effect observed in this case is suggested to be due to the reduced ability of the system to maintain a stable capillary neck as the sliding velocity increases.
On the other hand, in \cite{chen}, VW was associated with surfaces capable of forming hydrogen bond networks, in a process that proceeds according to some internal time scale.
Recently, Carpick {\em et al.}\cite{carpick} have found evidence 
that micro-contacts between solids can become chemically stronger as a function of time, an effect that can potentially produce VW.
In any case, VW is a manifestation of the existence of some internal dynamics of the contact, with its own time
scale. This time scale will ultimately determine the velocity range for which VW is observed.

To our knowledge ASs have not yet been observed in any nanofriction experiment. Actually, the only context where ASs are well characterized is that of seismic phenomena. Observation of ASs requires first of all of
a non-smooth sliding behavior, in such a way that relative displacement of the sliding materials occur discontinuously, 
and  a sufficiently low relative velocity in order to clearly separate the time scales of main shock (produced directly by the driving) and ASs. In the seismic context the discontinuous events are the earthquakes. In nanofriction these are the `slip' stages of a stick-slip process.\cite{persson} Yet, ASs are a very special type of abrupt sliding events. They are events that are caused by previous slip events, and that are unrelated directly to the driving. This means for instance, that if driving is stopped at some time $t_0$, there can still be some events occurring at $t>t_0$ which are aftershocks of those that occurred at $t<t_0$. Note that an unambiguous classification of an event as an aftershock requires a sufficiently slow sliding that does not trigger additional events in the mean time between the original event and its ASs. Also, it is clear from this description that ASs require the existence of some internal dynamics of the contact, in addition to the dynamics provided by the external driving.

The simplest model that describes a non-uniform, stick-slip dynamics in friction experiments is the Prandtl-Tomlinson (PT) model \cite{prandtl,tomlinson}, which we briefly review in the next section. In its simplest form, the PT model does not display VW.
In this paper we analyze two variants of the basic PT model that produce VW, and also ASs.
They describe two different implementations of relaxation mechanisms in the system. One can be applied (although not exclusively) to describe  a viscous damping in the tip of the AFM, and the second models the possibility of aging effects in the contact region of the sliding objects. 
We show clearly how ASs become progressively more defined as the relative sliding velocity decreases. 
In addition to the constant velocity setup, we also study the response to sudden changes in the driving velocity, looking for typical signatures in the transient behavior of the friction force such as a stress peak at sliding initiation, and a gradual relaxation of stress upon sliding stop.

The paper is organized as follows. In the next section we briefly review the basic phenomenology of the Prandtl-Tomlinson model that is relevant to our analysis, and discuss in detail the motivation of the two variants we will study. In Section III we show the results of the numerical simulations in a constant velocity set up, emphasizing the similarities and differences between the two, and showing in particular how ASs appear. In Section IV we give the results for some non-constant velocity protocols. Finally, In section V, we
summarize, and conclude.

\section{The Prandtl-Tomlinson model, and some variants} 

A basic understanding of the origin of friction in nanocontacts is provided by the Prandtl-Tomlinson (PT) model\cite{prandtl,tomlinson} in which the contact is represented by a single variable $x$ evolving under the effect of some interaction potential $W(x)$, that can be written as the product of a nominal distance between the two surfaces $h(x)$ and the normal force applied to the contact $F_N$, namely $W(x)\equiv h(x) F_N$. The variable $x$ is externally driven through an elastic coupling by a prescribed evolution $U(t)$.
The PT model adapts particularly well to the case of the AFM tip-substrate interaction, where  it is usually considered that $W(x)$ reflects the atomic corrugation potential. However, It has been used also to describe more general situations by using an appropriate disordered $W(x)$ function, and having in mind that the distance between typical minima of the $W(x)$ and the overall values of corrugation $h(x)$ are now defined by the roughness associated to the substrate.

\begin{figure}[h]
\includegraphics[width=8cm,clip=true]{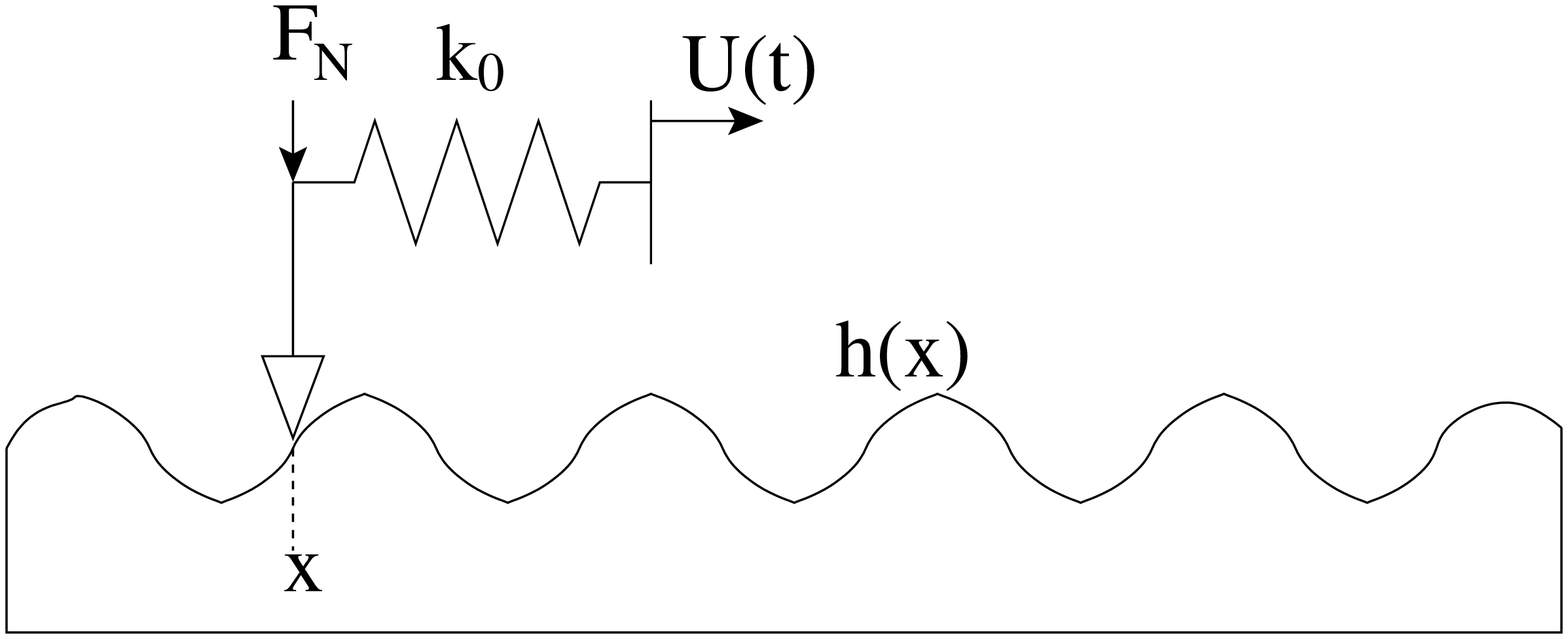}
\caption{The basic configuration of the Prandtl-Tomlinson model. The contact, represented by a single horizontal coordinate $x$, is pushed down with a force $F_N$ over a rough substrate defined by $h(x)$, and driven through a spring by an externally applied $U(t)$.
}
\label{sketch}
\end{figure}

In Fig. \ref{sketch} we sketch the basic configuration of the PT model. 
The evolution equation for $x(t)$ consists of a deterministic part associated to the forces applied to the tip, and a stochastic part originated in the integrated effect of many other ``rapid" degrees of freedom that are not explicitly taken into account in our description, which concentrates only in the ``slow" variable $x$. The effect of stochastic forces is twofold. On one hand they produce an additional term on the dynamic equation of the form of a white external noise with an intensity that is a function of temperature. On the other hand they are responsible for the appearance of ``friction" terms, namely additional forces that are proportional to minus the time derivative of the slow variables in the problem. 
By introducing these terms, we arrive at the basic equation describing the Prandtl-Tomlinson model\cite{nota}:
\begin{equation}
\eta_0 \dot x = (U-x)k_0 -\frac {dW}{dx} +\sqrt{2\eta_0 T}\phi(t)
\label{p-t}
\end{equation}
here $U(t)$ represents the driving, which for a constant velocity situation is $U(t)=Vt$, and $\phi(t)$ is the stochastic force associated to the fluctuating environment.
The instantaneous friction between the surfaces can be considered to be measured by the stretching of the $k_0$ spring, and the average friction force can be evaluated as the temporal average of this quantity, namely
\begin{equation}
F_{\mbox{fr}}=k_0 \overline {(U-x)}
\label{fuerza}
\end{equation}
We assume that the substrate potential term has some typical distance between minima $a$, and typical height $h_0$. 
We now give a brief overview of the phenomenology associated to this equation, and then discuss some possible variants that generate VW and other associated effects. 


We first describe the zero temperature case. The behavior of the system depends crucially on the ratio between $k_0$ and $d^2 W/dx^2$. 
For $k_0> \left . d^2 W/dx^2 \right |_{\mbox {max}}\simeq h_0 F_N/a^2$, there is a single equilibrium position of the tip for any value $U$ of the driving. This means that the dynamics is quasi-static for a vanishing driving velocity, and that the friction force vanishes in this limit. In the opposite case $k_0 < \left . d^2 W/dx^2 \right |_{\mbox {max}}$ there are metastable configurations of the tip, and abrupt transitions among them, as driving proceeds. This produces a finite friction force in the limit of driving velocity $V\to 0$. 
From now on, this is the case we consider.

At large driving velocities, the friction force becomes proportional to velocity, with a pre-factor given by the friction coefficient $\eta_0$. A crossover velocity that is order $\eta_0^{-1}$ can thus be defined, separating the high velocity regime of friction proportional to velocity, and a low velocity regime of rather constant friction force. 
We can formally describe this low velocity limit by taking  $\eta_0\to 0$ in Eq. (\ref{p-t}).

Temperature produces important changes in the velocity dependence of friction at very low velocities. In fact, at finite temperature, the system has the possibility to overcome the energy barriers before the pulling takes the system at the mechanical instability point. Since close to these instability points the energy barriers are much smaller for forward  than for backward movements, the thermal jumps always reduce the stretching of the $k_0$ spring, reducing the  friction force. 
The effect becomes more important as the velocity is reduced, leading typically to a logarithmic increase of friction force with velocity. A detailed account of this behavior can be found for instance in \cite{muser}.


Temperature effects are typically important in AFM measurements, at least at the lowest velocities, because of the small size and mass of the tip. They are less important when the size of the contact is larger, as in the surface force apparatus.
Here, in order to describe in the simplest possible form the effect of terms generating VW, we will work in the  $T=0$ limit. 
Also, as we are mostly interested in the limit of very small velocities, we will consider from now on the limit $\eta_0\to 0$ of Eq. \ref{p-t}.

\begin{figure}[h]
\includegraphics[width=8cm,clip=true]{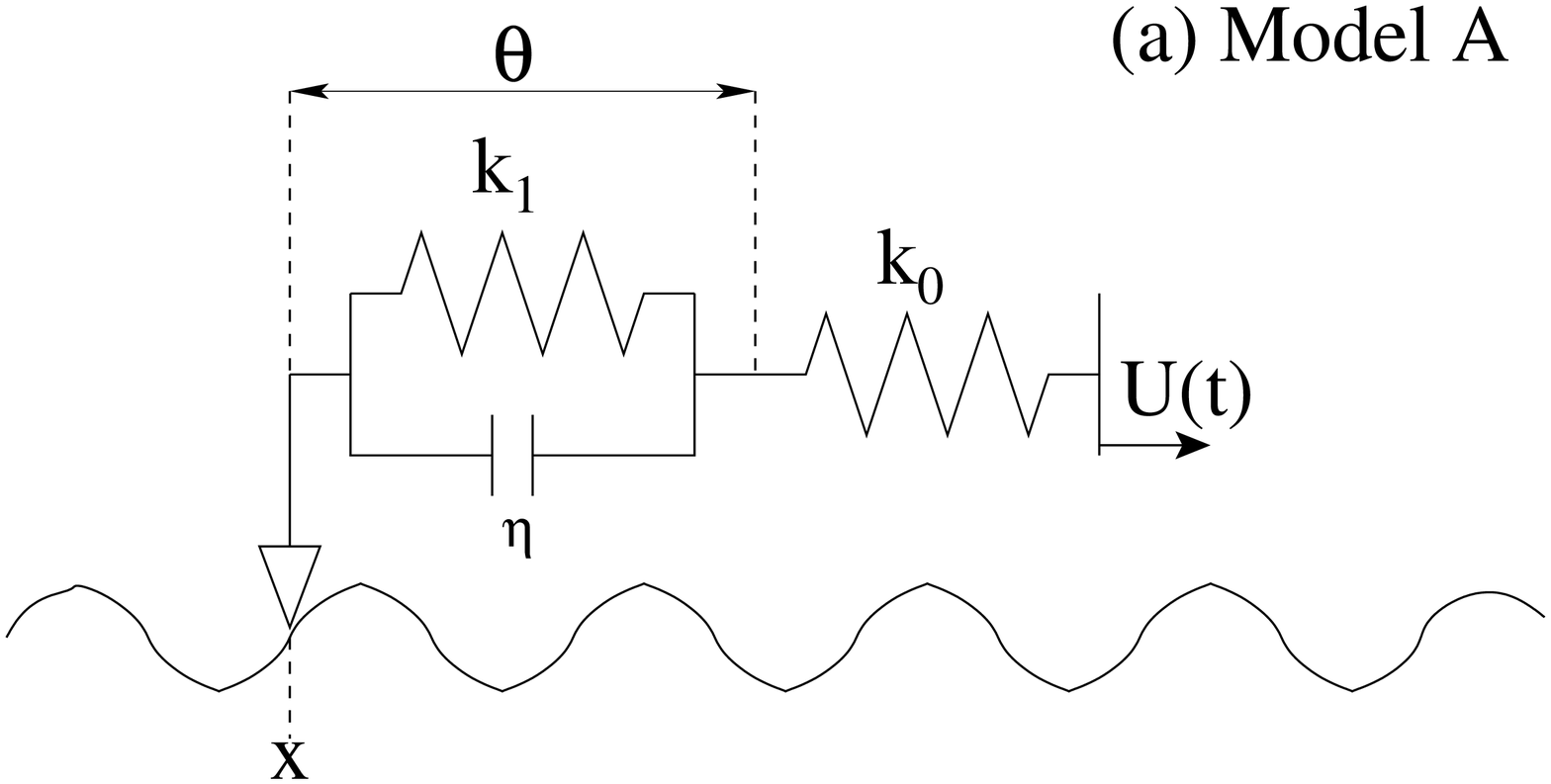}
\includegraphics[width=8cm,clip=true]{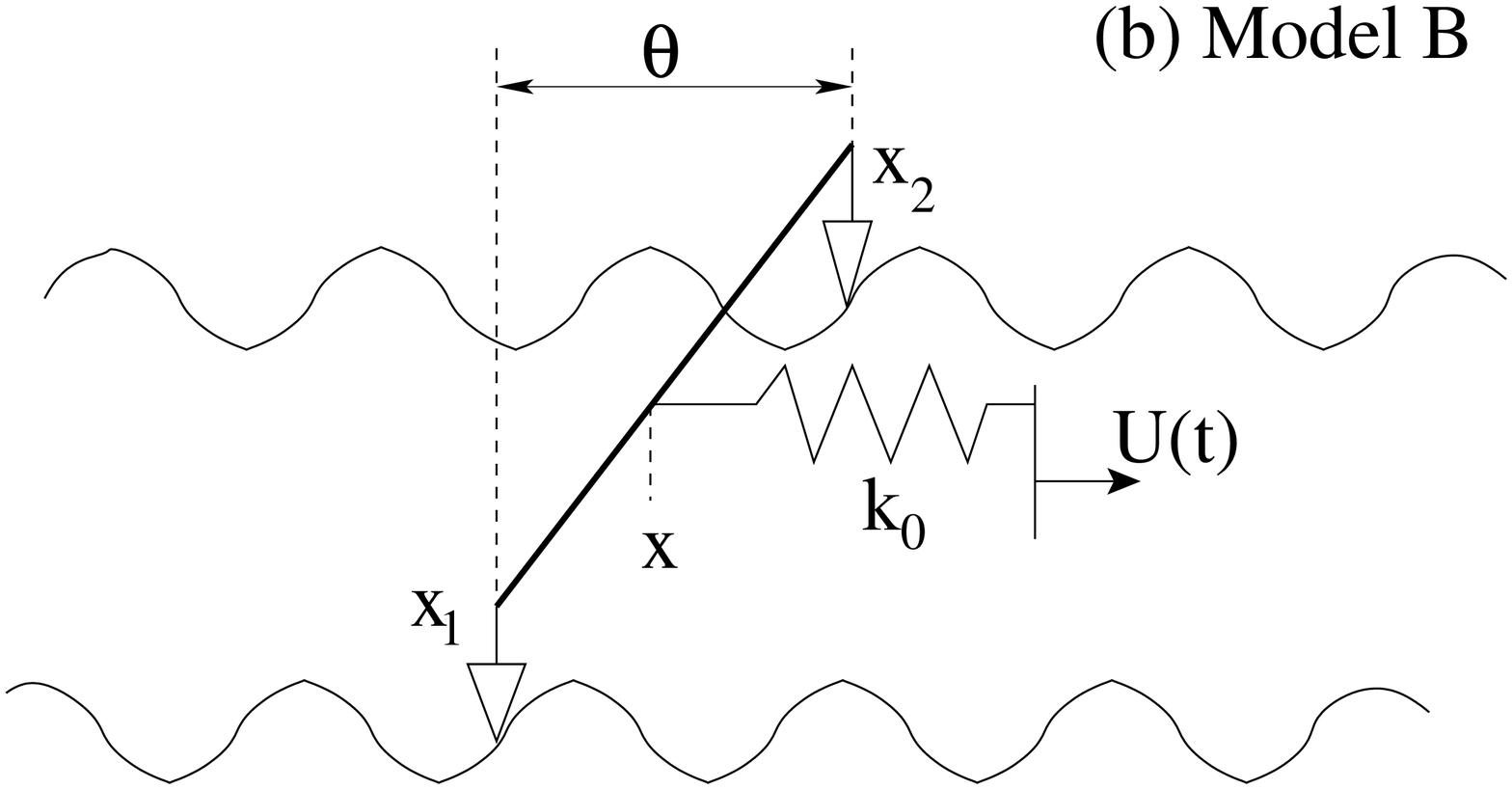}
\caption{Two variations on the basic configuration of Fig. \ref{sketch} that produce VW and ASs. In (a), the driving spring is complemented by a viscoelastic element. In (b) the tip is formally divided in two parts, each of them interacting with an independent rough substrate. The two parts are connected by a viscoelastic element (represented here by the thicker line). See the text for the equations defining both models.
}
\label{sketchab}
\end{figure}

We now introduce two different modifications to the Prandtl-Tomlinson model that produce VW. 
The first mechanism replaces the single spring $k_0$ driving the system, by a viscoelastic element as displayed in Fig. \ref{sketchab}(a). This produces a viscous damping when the system is driven, and introduces an additional degree of freedom, namely the variable $\theta$, and an internal time constant in the problem $\tau=(k_1+k_0)\eta$. In the particular case of the AFM setup, the present mechanism corresponds to incorporate viscous damping effects in the microscope tip. 
This is one example of the generalizations of the Prandtl-Tomlinson model known as the ``two masses-two springs" model\cite{krylov}. 
The full phenomenology of the two masses-two springs model is quite complicated, particularly if mass terms for the tip and cantilever are included. Here we will not introduce mass terms,
and the equations that describe this variant (that we call model A) read
\begin{eqnarray}
\eta_0 \dot x &=& (U-x-\theta)k_0 -\frac {dW}{dx} \nonumber\\
\eta\dot \theta &=&(U-x-\theta)k_0-\theta k_1
\label{a}
\end{eqnarray}

A similar kind of dissipation mechanism in the tip for AFM experiments has been discussed by 
Reimann and Evstigneev\cite{reimann1,reimann2}, who have shown how it can produce a VW effect. However, they did not present a detailed description of the sliding process, and in particular they do not refer to the possibility of ASs due to the viscous damping.

Model A describes a modification of the PT model in which the new internal dynamics, is introduced ``in series" with the contact, which itself is not modified.
A second possibility is to introduce relaxation right at the contact.
In order to do this, it seems necessary to go beyond the description of the contact in terms of a single slow variable $x$. 
In this respect we note that Evstigneev {\em et al.}\cite{evstigneev,evstigneev2} have provided evidence that in some cases the results of friction AFM experiments are incompatible with an interpretation in terms of a single variable characterizing the evolution. They have suggested that aging effects have to be included, and introduced a model with an additional degree of freedom describing a ``reinforcing" of the contact if this has lasted more than a certain time. 

Here we will make an effective description considering that the tip-substrate interaction region is not point like, but has some spatial extension. 
This will allow different parts of the contact region to accommodate relatively to each other to generate a stronger contact, if enough time is available.
A spatially extended tip-substrate contact forces us to consider several variables describing the interaction in different parts of this region. Trying to keep the model in the simplest non-trivial form, we will consider only two parts of the contact region, described by two variables $x_1$ and $x_2$. They will respectively move on disordered potentials $V_1$ and $V_2$ (producing forces $f_1$ and $f_2$) which for simplicity will be assumed to be uncorrelated. 
The average position of the tip is defined as $x\equiv (x_1+x_2)/2$. The relaxation effects will be modeled through a viscoelastic response of the variable $\theta \equiv x_1-x_2$ that defines the internal state of the contact. 
A pictorial sketch of the model, called model B, is shown in Fig. \ref{sketchab}(b). The evolution equations for $x$ and $\theta$, that completely define the dynamics of this model are as follows

\begin{eqnarray}
\eta_0 \dot x&=&(U-x) k_0 +f_1(x+\theta/2)+f_2(x-\theta/2)\nonumber\\
(\eta+\eta_0) \dot \theta&=&f_1(x+\theta/2)-f_2(x-\theta/2) -k_1 \theta
\label{b}
\end{eqnarray}
When applied to AFM experiments, where there is an obvious asymmetry between the two sliding objects, the presentation we made of model B suggests that we are introducing a relaxation mechanism in the tip, and that substrate is inert. This is not necessarily so, as alternative presentations with essentially the same final equations can be produced in which the relaxation effect is mainly concentrated in the substrate, or in both. For concreteness, we refer only to the previous realization.

\section{Constant velocity results}

In this section we solve numerically Eqs. (\ref{a}) and (\ref{b}) for models A and B, at constant values of the driving velocity and in the limit of $\eta_0\to 0$. This is achieved through the following protocol. We set the position of the driving at some constant value $U$, and iterate the evolution equations using a simple first order algorithm, until an equilibrium position is obtained for $x$ (i.e, until $\dot x$ is sufficiently close to zero). This process is assumed to occur at fixed time, as $\eta_0\to 0$. After this, time is advanced a small $dt$, which produces an increase of $U \to U +Vdt$, and the process is repeated. 

The disordered pinning potential we use is obtained through the following interpolation mechanism. For all points of the horizontal axis with coordinates of the form $x_i\equiv a i$, with $i$ an integer, a Gaussian random number $\phi_i$ of zero mean and unitary variance is generated.  Each $\phi_i$ produces a pinning potential that is zero outside the interval $[x_{i-2},x_{i+2}]$, and has the form $W_i(x)=h_0 \phi_i \left [\cos\left (\pi (x-x_i)/2a\right ) +1\right ]$ inside it. The total $W(x)$ is the sum of the individual contributions: $W(x)=\sum_i W_i(x)$. Note that there are at most four non zero terms in this sum.
In the case of model B, we scale down the forces $f_1$ and $f_2$ by a factor $\sqrt{2}$, in order to have typically the same overall friction force as in model A.
Also, we will present all results in dimensionless form, using $a$ as the unit of length, $\eta a^2/F_N h_0$ as the unit of time, and $F_N h_0/a $ as the unit of force.

\begin{figure}[h]
\includegraphics[width=8cm,clip=true]{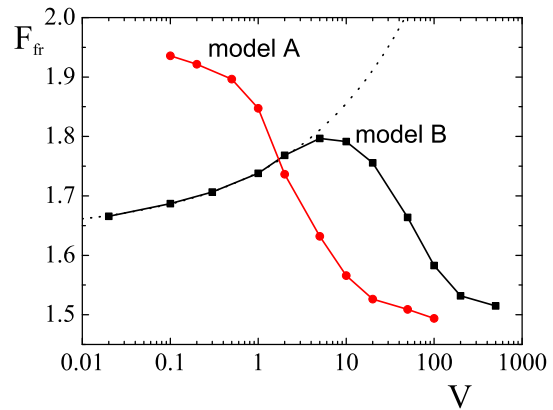}
\caption{Average friction force as a function of the pulling velocity, for the two models sketched in the previous figure. Both of them display velocity weakening (results here and in the following figures were obtained for $k_0=k_1=0.3$). Model $V$ also shows a velocity increase of friction force at low velocities that follows an $\alpha+\beta V^{1/3}$ law (dotted line).
}
\label{vdet}
\end{figure}

The results for the average friction force as a function of velocity are contained in Fig. \ref{vdet}. The two models clearly display velocity weakening in some velocity range (model B also displays increase of friction force at the lowest velocities, we address this effect later on). The qualitative justification for the existence of VW applies to the two models in the same way, and goes as follows. At large velocities, 
the driving induces temporal variations in the dynamics with a period $T=a/V$. On the other hand, 
the dissipative element possess a time constant $\tau$ given by $\tau= \eta \tilde k$ ($\tilde k= k_0+k_1$ for model A, and $\tilde k= k_1$ for model B).
For $\tau \gg T$ the $k_1$ spring is effectively blocked by the dash-pot, namely $\theta$  is nearly constant for sufficiently large velocities. On the contrary, at the lowest velocities, there is plenty of time for $\theta$ to adapt, which generates an additional dissipation that manifests in  an increase of the friction force. 

\begin{figure}[h]
\includegraphics[width=8cm,clip=true]{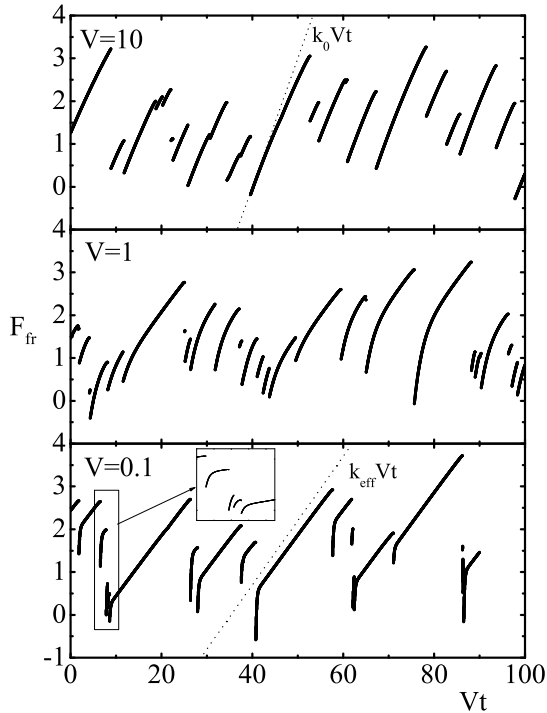}
\caption{Instantaneous friction force as a function of time, for model A (see Fig. \ref{sketchab}(a)), at different pulling velocities. At large velocity (upper panel) the model behaves almost as a normal PT model with spring constant $k_0$. As velocity is reduced, there is an additional response associated to the dissipative element $\eta$ and the $k_1$ spring. At the lowest velocities (lower panel), and except for time intervals of order $\tau\equiv (k_0+k_1)\eta$ after the force jumps, the response is equivalent to a normal PT model with a pulling spring of effective constant $k_{eff}\equiv k_0k_1/(k_0+k_1)$. In time intervals of order $\tau$ after the jumps, ASs can be observed. One example is highlighted.}
\label{xdet2}
\end{figure}

The time evolution of the instantaneous friction force reveals some differences between models A and B. Let us start by describing the evolution for model A. The instantaneous friction force at different velocities is shown in Fig. \ref{xdet2}.
As previously stated, the evolution for large velocities corresponds to that in the original PT model with a driving spring $k_0$. We observe the typical jumps in the friction force, corresponding to rapid rearrangement of the $x$ variable, that is the origin of a finite average force. 
At the lowest velocities, we see that after the jumps, there is a rapid recovering of the force, which, with a time constant given by $\tau= (k_0+k_1)\eta$ approaches a linear behavior in which force increases with a spring constant corresponding to an effective stiffness $k_{eff}\equiv k_0k_1/(k_0+k_1)$. So there is a transition in the stick stage of the movement, passing from  an effective constant $k_0$ at large velocities, to $k_{eff}$ at low velocities. 

\begin{figure}[h]
\includegraphics[width=8cm,clip=true]{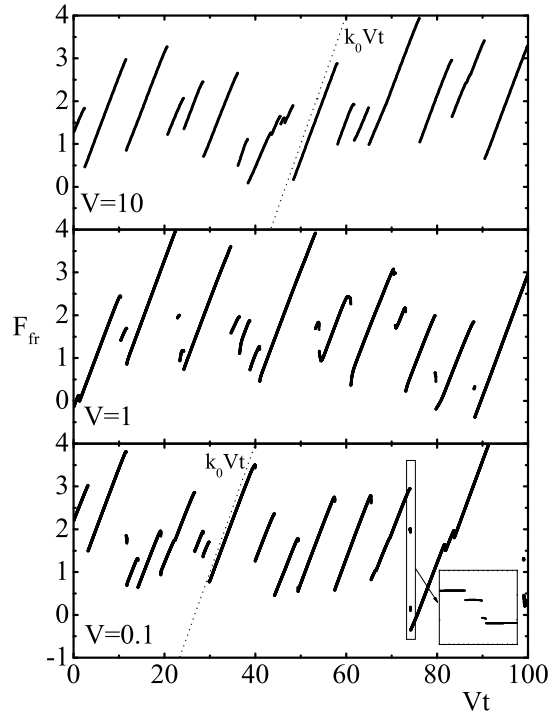}
\caption{Same as previous figure for model B (see Fig. \ref{sketchab}(b)). Although this model also shows VW and ASs, the slope of the force increase during the `stick' stage of the movement corresponds always to a stiffness $k_0$. 
}
\label{xdet4}
\end{figure}

In the case of model B (Fig. \ref{xdet4}), although velocity weakening is equally observed, the time evolution of instantaneous force does not display a transition between two different spring constants, instead, it always corresponds to an evolution with stiffness $k_0$. This difference in the time evolution of friction force can be a potentially useful tool to discern between the different mechanisms producing VW in a particular experimental realization.


\begin{figure}[h]
\includegraphics[width=8cm,clip=true]{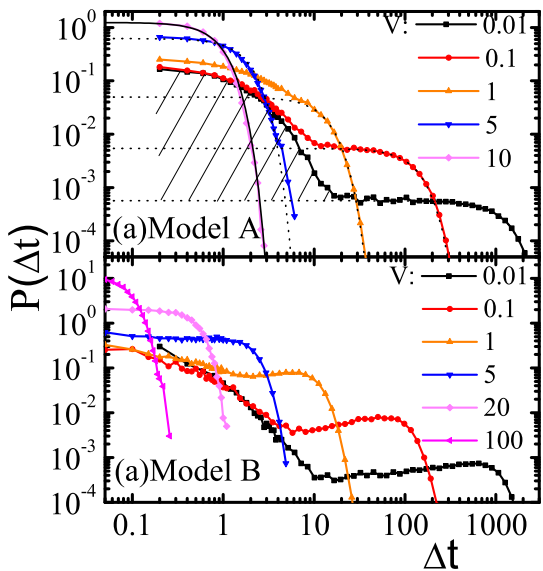}
\caption{(a) Histograms of time intervals between jumps for model A  at different driving velocities. Al large velocities, the distribution is well fitted by a Gaussian distribution (continuous line). When this distribution is rescaled to be applied to lower velocities (dotted lines) it only fits the large $\Delta t$ part of the distributions. The additional events with low $\Delta t$  are ASs (for the lowest velocity, ASs are indicated by the dashed region). Note that ASs are not affected by $V$, once $V$ is small enough to allow their observation.
(b) The same qualitative behavior is observed for model B, although with slightly different forms of the distributions.
}
\label{aftershs}
\end{figure}

In addition to display VW, both A and B models also possess aftershocks, as we will discuss now. Let us concentrate for concreteness in model A. 
The jumps in the evolution of the friction force correspond to transitions between different metastable states of the system. As it was already discussed, and it is clear in Fig. \ref{xdet2}, at very low velocities the instantaneous friction force after a jump displays a dip, and then a recovery during a typical time $\tau$. If this force increase is enough to destabilize the new position in which $x$ is located, this will produce a second jump to a more stable minimum. This second jump is an aftershock to the first one. Note that aftershocks (that can in fact be more than one for a single initial jump) occur in a time scale of order $\tau$ which depends exclusively on the internal dynamics of the system.
In fact, at sufficiently low velocities, it is observed that sometimes there exist sequences of a few jumps that occur close in time (actually, within a time interval of order $\tau$) and they represent the initial shock and its aftershocks. Some particular examples are highlighted in Figs. \ref{xdet2} and \ref{xdet4}.
A more quantitative description of ASs can be made by generating the histogram of time intervals between consecutive jumps in the instantaneous friction force. 
The typical time interval expected between main jumps is given by $\sim a/V$. On the other hand, the typical time interval for the appearance of an aftershock is $\tau$. It is clear that in order to unambiguously identify ASs we must have $V \ll a/\tau$.  This is in fact what is observed in the results shown in Fig. \ref{aftershs}(a). At large velocities, we see a smooth distribution of time intervals that for the present form of the pinning potential can be rather well approximated by a Gaussian distribution. In the absence of relaxation (i.e., $\eta\to \infty$), as velocity is reduced we would simply obtain the same distribution but stretched to larger time intervals. These rescaled curves are shown in Fig. \ref{aftershs}(a) as continuous lines. We see that for sufficiently low velocities, the true distribution of time intervals displays a peak at small time intervals of order $\tau$ (independent of $V$). This is the peak corresponding to ASs. For the parameters used in Fig. \ref{aftershs}(a) we find that about one quarter of all jumps are ASs.  We emphasize that ASs are visible only at sufficiently low velocities, otherwise they are masked by the jumps that are produced by the driving itself.

The situation for model B is quite similar. The histograms of time intervals (shown in Fig. \ref{aftershs}(b)) display the same features, although the actual form of the distributions are slightly different. 
In the case of model B, ASs occur when in the process of relaxing the variable $\theta$, the system finds an instability point for the center of mass variable $x$. 
From Fig. \ref{aftershs} we can conclude that the average time between main shocks $a/V$ has to be larger than about 10 in order than ASs can be identified. Restoring units, this gives $V\lesssim 0.1 F_N h_0 /\eta a$ as necessary to observe ASs.

Model B presents at the lowest velocities a region of velocity strengthening (see Fig. \ref{vdet}), namely , an increase of friction force upon velocity increase. A detailed analysis of the time evolution of $x$ and $\theta$ reveals the mechanism that originates the effect. 
Consider the limit $V\to 0$, and imagine the trajectory of the coordinates of the system in the $x$-$\theta$ plane. The trajectory consists of smooth pieces, and abrupt transitions when instabilities occur. At a somewhat larger value of $V$, the trajectory follows basically the same path that in the $V\to 0$ limit, but the abrupt transitions between smooth pieces of trajectories is slightly delay due to the time constant of the $\theta$ variable. This produces an increase in friction force through the same mechanism that operates in Eq. (\ref{p-t}) for non-zero $\eta_0$. In that case an increase of friction force as $\sim V^{2/3}$ is well known \cite{muser,fisher}, and originates in the delay that the $x$ variable experiences in trying to follow the quasistatic behavior, when $\eta >0$. In the present case, it is the $\theta$ variable that takes some time to adapt to the evolving landscape. In this case however, the instability occurs through a Hopf bifurcation, and this modifies the velocity dependence, which now is of the form $V^{1/3}$.
In fact, the low velocity increase of the friction force is very well fitted by this velocity dependence (see Fig. \ref{vdet}).

\section{Results for slide-stop-slide experiments}

\begin{figure}[h]
\includegraphics[width=8cm,clip=true]{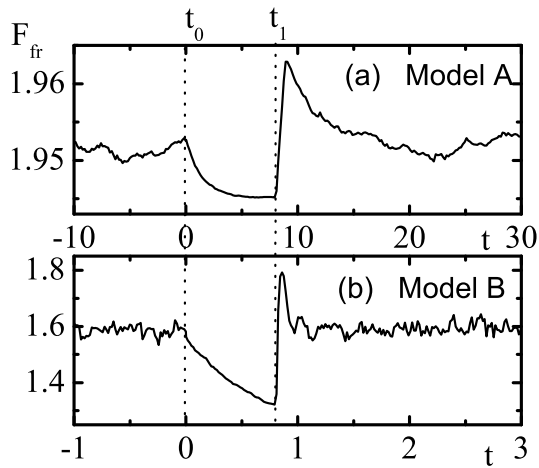}
\caption{Average friction force over many realizations for a system that is initially sliding at velocity $V$. Sliding stops at $t=t_1$ and is reinitiated at the same $v$ at $t=t_2$. Results are for model A [(a) $V=0.1$, 2000 realizations] and [(b), $V=100$,  $10^5$ realizations] with other parameters as in previous figures.
}
\label{v1v2}
\end{figure}

The mechanisms that lead to VW in geological materials produce also characteristic features in experiments in which the velocity of the driving is 
changed in some abrupt way. For instance, if driving is stopped, there is typically a logarithmic decrease in time of the stress in the system, as the internal degrees of freedom relax. When driving is reinitiated, there is a transient in which the stress displays an overshoot with respect to the stationary value, which reflects the higher energy barrier that has to be overcome to reinitiate sliding, after a period of rest. It is interesting to see if this phenomenology is also observed in the simple generalizations of the PT model we are discussing here. 
Due to the large variations of the instantaneous friction force in our system, it is clear that a systematic effect will only be observable if an average over many realizations is performed.
In Fig. \ref{v1v2} we show this average for the instantaneous friction force as a function of time in a system that is initially driven at a fixed velocity. The driving is stopped at time $t_0$, and re-initiated at time $t_1$ with the same original velocity. The results show that the stress reduction in time after sliding stops, and the peak after reinitiation, are
qualitatively reproduced. The detailed form of the time dependencies however, cannot be expected to match the experimental ones \cite{marone,marone2}, as they depend on collective effects that are not captured by the simple models discussed here.

\section{Summary and conclusions} 

Velocity weakening and aftershocks are two key ingredients that characterize the dynamics of seismic phenomena. They are deeply related with the existence of relaxation effects within the sliding materials.  When going progressively to smaller spatial scales, these two phenomena have been difficult to observe. 

We have shown however that some simple relaxation mechanisms, that may reasonable be acting in microcontact experiments, can lead to the expectation that these phenomena may occur and be observable in a variety of situations. VW and ASs were obtained here by introducing two different types of relaxation in the standard PT model. In one of them, the driving spring shows a viscoelastic response. In the second, the contact region was assumed to have some spatial extent and to be described by more than a single variable. In both cases, there is an internal time constant that characterizes the response of the system. VW and ASs are observed when the driving velocity leaves enough time to the contact to adapt its internal degree of freedom as sliding proceeds.

Temperature effects have not been consider in our analysis. It is clear that thermal fluctuations will be a competing effect against the observation of VW ans ASs. As thermal fluctuation effect are more important in smaller size devices, the observation of ASs with the AFM may be a challenging task. 

The two model analyzed here were constructed 
having in mind situations of very small contact regions, like those realized in the surface force apparatus, or even at smaller scales, with the AFM. Applications to larger contact regions likely require the consideration of an extended model with many degrees of freedom. In this respect the consideration of a model similar to the present model B in a situation with many contact point has been considered for instance in  \cite{jagla1,jagla2}, and in fact led to a phenomenology comparable with that observed in earthquakes. The consideration of model A in a similar context has not been done yet, and remains as a future prospect.

\section{Acknowledgments}

This research was financially supported by Consejo Nacional de Investigaciones Cient\'{\i}ficas y T\'ecnicas (CONICET), Argentina. Partial support from grant PIP112/200901/51 (CONICET, Argentina) is also acknowledged.


\begin{thebibliography}{4}

\bibitem{carpick0}I. Szlufarska, M. Chandross, R. W. Carpick,  J. Phys. D: Appl. Phys.
{\bf 41}, 123001 (2008).


\bibitem{sfa}J. Israelachvili {\em et al.} Rep. Prog. Phys. {\bf 73} 036601 (2010).

\bibitem{scholz}C. H. Scholz, {\em The Mechanics of Earthquakes and Faulting},  (Cambridge University Press, Cambridge, England, 2002).

\bibitem{dietrich}J. H. Dieterich, J. Geophys. Res. {\bf 84}, 2161 (1979).

\bibitem{ruina}A. Ruina, J. Geophys. Res. {\bf 88}, 10359 (1983).

\bibitem{rs_and_as}J. H. Dieterich, J. Geophys. Res. {\bf 99}, 2601 (1994).

\bibitem{jagla1}E. A. Jagla and A. B. Kolton, J. Geophys. Res. {\bf 115}, B05312 (2010).

\bibitem{jagla2}E. A. Jagla, Phys. Rev E {\bf 81}, 046117 (2010).

\bibitem{marone}C. Marone, Nature {\bf 391}, 69 (1998).

\bibitem{marone2}C. Marone, Annu. Rev. Earth Planet. Sci. {\bf 26}, 643 (1998).

\bibitem{riedo}E. Riedo, F. L\'evy, and H. Brune, Phys. Rev. Lett. {\bf 88}, 185505 (2002).

\bibitem{chen} J. Chen, I. Ratera, J. Y. Park, and M. Salmeron, Phys. Rev. Lett. {\bf 96}, 236102 (2006).

\bibitem{carpick}Q. Li, T. E. Tullis, D. Goldsby, and R. W. Carpick, Nature {\bf 480}, 7376 (2011).


\bibitem{persson}B. N. J. Persson, {\it Sliding Friction, Physical Principles and Applications}, (Springer, Berlin, 2000).

\bibitem{prandtl}  L. Prandtl, Z. Angew. Math. Mech. {\bf 8}, 85 (1928).

\bibitem{tomlinson} G. A. Tomlinson, Philos. Mag. {\bf 7}, 905 (1929).

\bibitem{nota}We will not  discuss inertia effects, namely, the existence of a second time derivative term in this equation.
This term is known to produce VW, however in a large velocity regime \cite{muser}, which is not our main interest here.



\bibitem{muser}M. H. Muser,  Phys. Rev. B {\bf 84}, 125419 (2011).

\bibitem{krylov}S. Y. Krylov and J. W. M. Frenken, J. Phys. Condens. Matter {\bf 20}, 354003 (2008). 

\bibitem{reimann1}P. Reimann and M. Evstigneev, Phys. Rev. Lett. {\bf 93}, 230802 (2004). 

\bibitem{reimann2}P. Reimann and M. Evstigneev, New J. Phys. {\bf 7},  25 (2005).

\bibitem{evstigneev}M. Evstigneev, A. Schirmeisen, L. Jansen, H. Fuchs, and P. Reimann, Phys. Rev. Lett. {\bf 97}, 240601 (2006). 

\bibitem{evstigneev2}M. Evstigneev, A. Schirmeisen, L. Jansen, H Fuchs, and P. Reimann,  J. Phys.: Condens. Matter {\bf 20}, 354001 (2008).

\bibitem{fisher} D. S. Fisher, Phys. Rev. B 31, 1396 (1985).

\end{thebibliography}
\end{document}